\documentclass[aps,prl,twocolumn,a4paper,superscriptaddress,intlimits,floats,floatfix,showpacs]{revtex4-1}
\newcommand\ho{SrHo$_2$O$_4$}
\newcommand\dy{SrDy$_2$O$_4$}

\usepackage{graphicx}
\usepackage[version=3]{mhchem}
\usepackage{color}

\begin{document}


\title{Evidence for \ho ~and \dy\ as model $J_1$-$J_2$ zig-zag chain materials }



\author{A. Poole}
\email{amy.poole@psi.ch}
\author{V. Y. Pomjakushin}
\affiliation{Laboratory for Neutron Scattering, Paul Scherrer Institute, CH-5232 Villigen, Switzerland}

\author{A. Uldry}
\author{B. Delley}
\affiliation{Condensed Matter Theory Group, Paul Scherrer Institut, CH-5232 Villigen PSI, Switzerland}

\author{B.~Pr\'{e}vost}
\author{A.~D\'{e}silets-Benoit}
\author{A.~D.~Bianchi}
\affiliation{ D\'{e}partement de Physique \& Regroupement Qu\'{e}b\'{e}cois sur les Mat\'{e}riaux de Pointe (RQMP), Universit\'{e} de Montr\'{e}al, Montr\'{e}al, Qu\'{e}bec H3C 3J7, Canada}

\author{R.~I.~Bewley}
\affiliation{ISIS Facility, Rutherford Appleton Laboratory, Chilton, Didcot, OX11 0QX, United Kingdom}

\author{B. R. Hansen}
\affiliation{Department of Physics, Technical University of Denmark, 2800 Lyngby, Denmark}

\author{N.~Kurita}
\author{R.~Movshovich}
\author{T.~Klimszuk}
\author{F.~Ronning}
\affiliation{ Condensed Matter and Thermal Physics, Los Alamos National Laboratory, Los Alamos, NM 87545, USA}

\author{R.~J.~Cava}
\affiliation{ Department of Chemistry and Princeton Materials Institute, Princeton University, Princeton, NJ 08544, USA}

\author{M.~Kenzelmann}
\affiliation{ Laboratory for Development and Methods, Paul Scherrer Institut, CH-5232 Villigen PSI, Switzerland}%


\date{\today}

\begin{abstract}
Neutron diffraction and inelastic spectroscopy is used to characterize the magnetic Hamiltonian of \ho\ and \dy.  Through a detailed computation of the crystal-field levels we find site-dependent anisotropic single-ion magnetism in both materials and diffraction measurements show the presence of strong one-dimensional spin correlations.  Our measurements indicate that competing interactions of the zig-zag chain, combined with frustrated interchain interactions, play a crucial role in stabilizing spin-liquid type correlations in this series.
\end{abstract}


\maketitle

Geometrically frustrated magnetic materials have proven to be a fertile area of condensed matter research.  Competition between interactions can lead to macroscopic degeneracies and novel states of matter with emergent properties, providing `toy models' for statistical mechanics and examples of exotic quasi-particle excitations.  A prime example are the rare-earth titanates in which the combination of the rare-earth ion dependent crystal field anisotropy and a highly frustrated lattice produce a rich array of novel magnetic materials with both classical and quantum spin liquid phases \cite{Gardner:2010p2400,Greedan:2006p670}.

\begin{figure}
   \centering
 \includegraphics[scale=0.35]{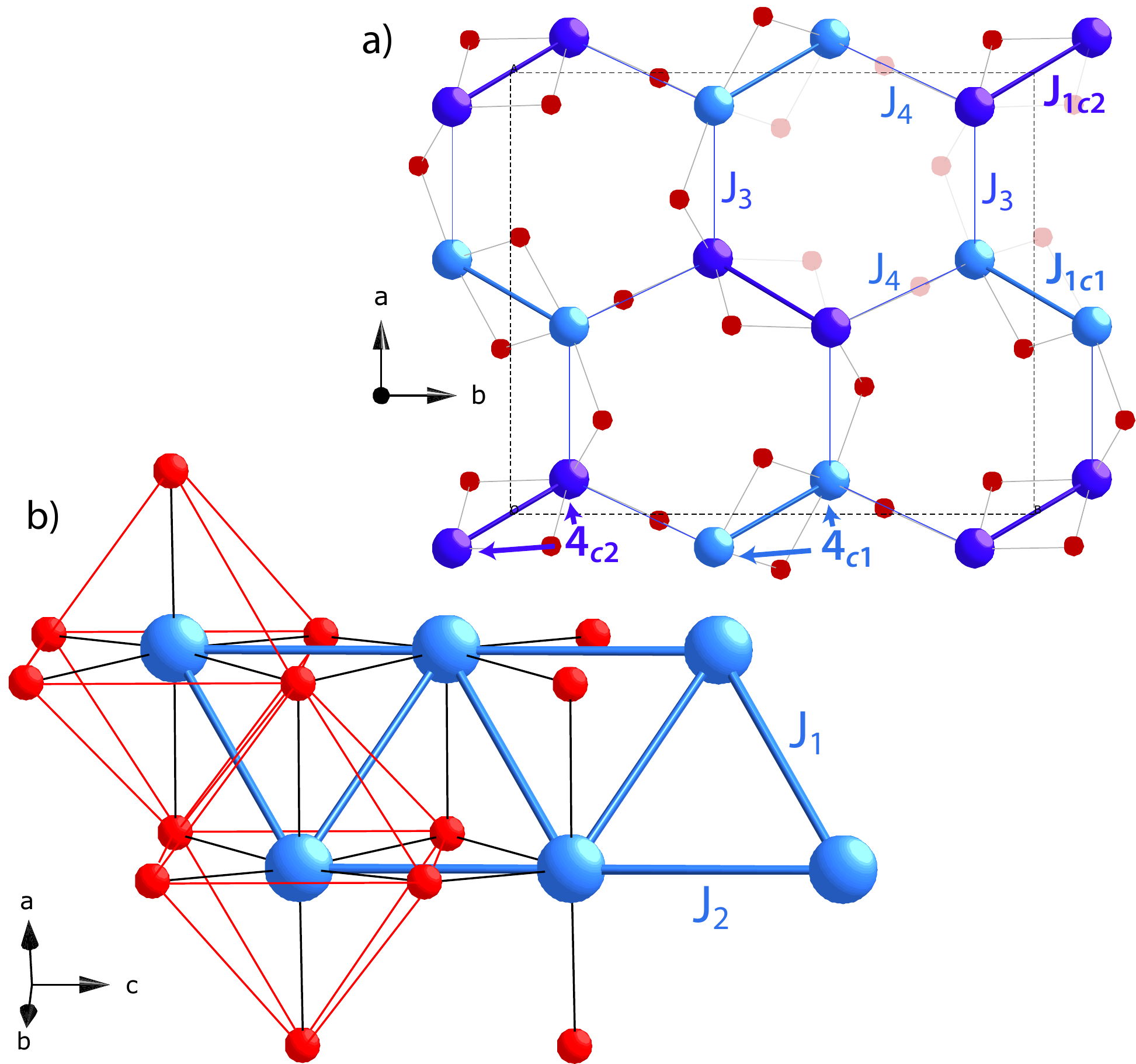}
   \caption{The structure of \dy.   The $4c_1$ and $4c_2$  sites are dark and light blue and oxygen is shown in red, respectively, a) showing the hexagonal arrangement of the atoms in the $ab$ plane, b) the Dy$_{4c1}$ ladder structure with the $J_1$-$J_2$ exchange pathways.  \cite{Karunadasa:2005p3755} }
   \label{fig:structure}
\end{figure}

\ce{Sr$R$2O4} belongs to a recently discovered family of geometrically frustrated rare-earth materials that features non-trivial ground states \cite{Karunadasa:2005p3755, Petrenko:2008p6833, Hayes:2012p6555, QuinteroCastro:2012p7953,Young:2013p7975}.   The rare earth sites ($R$) form a honeycomb in the $ab$ plane (Fig. \ref{fig:structure}.a)  and a triangular ladder along the $c$-axis (Fig. \ref{fig:structure}.b).  \ho\ was shown to have a one-dimensionally correlated state at low temperatures, with moments that lie along either the $b$ or $c$ axes \cite{Young:2013p7975}, whereas \dy\ shows only short-range order and weak diffuse scattering \cite{Karunadasa:2005p3755}.  In this Letter, we will demonstrate: firstly, that the two rare-earth sites feature a strong anisotropy pointing along the $b$ or $c$ axes, respectively; secondly, that \dy\ features one-dimensional correlations with up-up-down-down local order, but remains disordered on long length scales to the lowest measured temperatures.  We argue that the magnetism in \ce{Sr$R$2O4} can be mapped on to the Ising $J_1$-$J_2$ spin chain model, and that the competing interactions of this model play a crucial role in the stabilization of the spin liquid state in \dy\ \cite{Majumdar:1969p7877, Morita:1972p7901, HeidrichMeisner:2007p7967}.

The relation between the ladder structure and the $J_1$-$J_2$ model can be understood from Fig. \ref{fig:structure}, where the rungs and legs correspond to the $J_1$ and $J_2$ interactions, respectively.  Theory predicts that in $S=\frac{1}{2}$ systems with antiferromagnetic $J_1$, when $J_2> J_1/2$, the ground state changes from a simple antiferromagnetic N\'eel, $\uparrow \downarrow \uparrow \downarrow$, state to an up-up-down-down, $\uparrow \uparrow \downarrow \downarrow$, double N\'eel configuration \cite{Igarashi:1989p7867, Majumdar:1969p7877}.   On the application of a magnetic field the system enters an up-up-down phase, seen as a $\frac{1}{3}$ magnetisation plateau at H = -$J_1/2+J_2$ and finally saturates into a ferromagnetic phase, when H = $J_1+J_2$, in which all of the moments are aligned with the applied magnetic field \cite{Morita:1972p7901,Igarashi:1989p7867,Okunishi:2003p7870}.  A $\frac{1}{3}$-plateau is also found for classical moments, when the moments are strongly Ising like \cite{HeidrichMeisner:2007p7967}.   Despite several examples of Heisenberg $J_1$-$J_2$ \cite{Castilla:1995p7939, Enderle:2005p7932, Drechsler:2007p7935}, and $J_1$ Ising \cite{Goff:1995p7928, Wolf:2000p7944, Coldea:2010p7931}, materials there are few previous examples that meet the criteria for the Ising $J_1$-$J_2$ chain model \cite{Matsuda:1995p7933}, and none that can test the phase diagram at the classical limit.

The \ce{Sr$R$2O4} crystallographic structure is described by the $Pnam$ space group, and each unit cell contains a total of eight $R$ atoms that are divided into two inequivalent 4$c$ sites (4$c_{1}$ and 4$c_{2}$) at the center of distorted, edge-sharing, oxygen octahedra (Fig. \ref{fig:structure}) \cite{Karunadasa:2005p3755}.  The  4$c_{1}$ and 4$c_{2}$ sites form separate zig-zag chains, with rung exchanges $J_{1_{c1}}$ and $J_{1_{c2}}$, and leg exchanges $J_{2_{c1}}$ and  $J_{2_{c2}}$.   The $R$ ions have a monoclinic site symmetry and the free ion ground state of the $R$ atoms is split into the maximum  possible number of levels.  For the $^5I_{8}$, Ho$^{3+}$ ion, with an integer value for $J$, we expect $2J+1=17$ singlets and for the $^6H_{15/2}$, Dy$^{3+}$, with a half integer value for $J$,  the site symmetry gives rise to $J+1/2=8$ doublets \cite{Walter:1984p7623}.  Magnetisation studies suggest that each of the sites, in both the Ho and Dy materials, have large single-ion anisotropies, with  $b$ and $c$ easy-axis directions \cite{Hayes:2012p6555}.  The measured entropy indicates that the Dy moments have S = $\frac{1}{2}$ degrees of freedom \cite{Cheffings:2013p7898}.

\begin{figure}
\includegraphics[width=8.5cm]{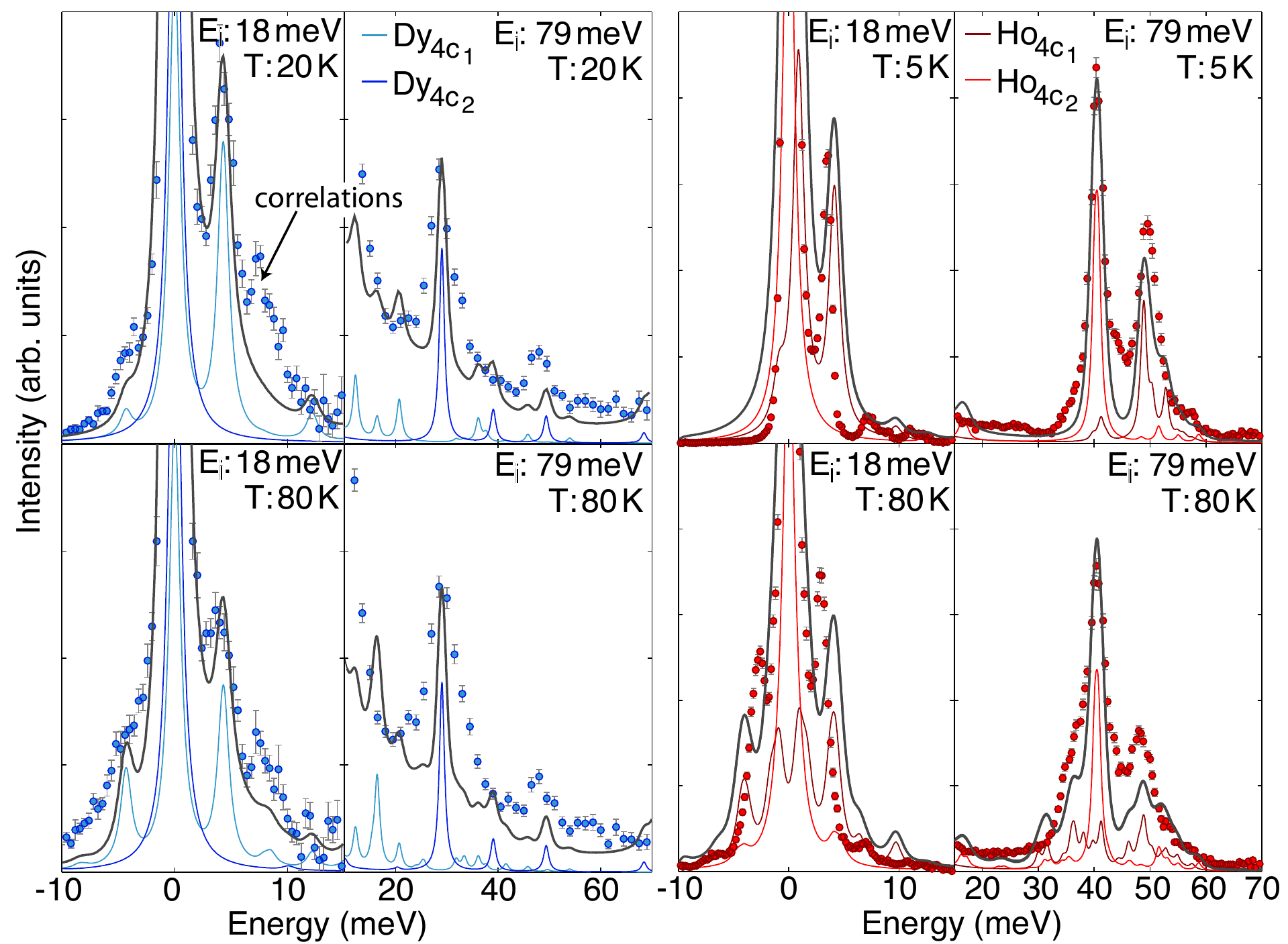}
\caption{\label{fig:cef}
(Color online) Inelastic neutron powder data for \dy\ (blue, left) and \ho\ (red, right) measured on HET.  The calculated fits for the individual sites are shown with thin solid lines and their average is indicated with the thick grey line.  The average has been convoluted with the instrument resolution and an instrumental background has been added to the Dy data.}
\end{figure}

To determine the crystal field excitations and determine the magnetic single-ion anisotropies, inelastic neutron scattering experiments were performed on powder samples of \dy\ and \ho\,  prepared by the method described in Ref.~\cite{Karunadasa:2005p3755}.  The inelastic neutron spectra of the Dy compound were recorded with an incident energy E$_i$=18 meV, 79 meV and 117 meV at temperatures of T=20 K, 50 K and 80 K, on the HET time-of-flight spectrometer, ISIS.  The Ho compound was measured with an incident energy of  E$_i$=18 meV, 70 meV and 79 meV at temperatures of T=5 K, 50 K, 80 K, and 150 K on the HET spectrometer.  Low energy \ho\ spectra were further measured on the time-of-flight spectrometer FOCUS at PSI at dilution temperatures.  The resolution of the spectrometers was determined from fits to the elastic line.  The instrumental background of the Dy data was found from a polynomial fit to the 117 meV data, which was then scaled by the incident energy.

\begin{figure}
\includegraphics[width=6.5 cm]{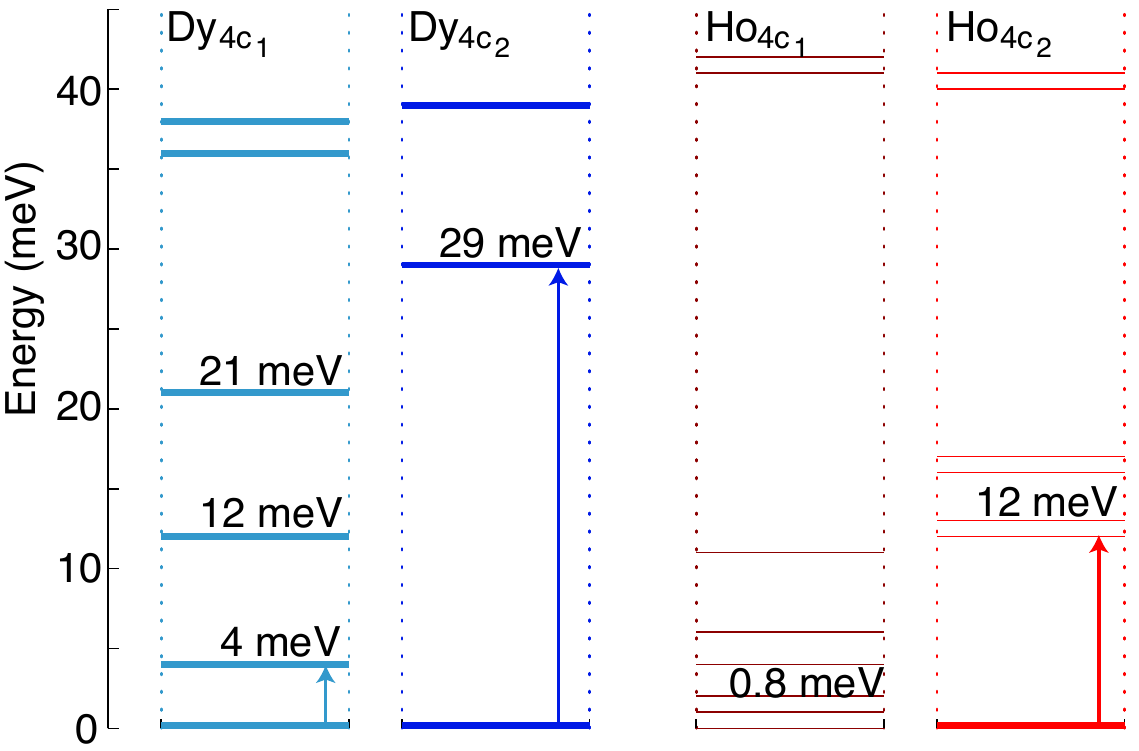}
\caption{\label{fig:ceflev}
(Color online) The level scheme of \dy\ (blue, left) and \ho\ (red, right).  The thick lines indicate a doublet state. The ground state of Ho$_2$ is a pseudo-doublet.}
\end{figure}

Due to the low symmetry of ion sites the excitations were modeled using a point charge calculation \cite{Uldry:2012p7902}.  These calculations capture the intra-atomic electrostatic interactions, the spin-orbit coupling and the effect of the crystal electric field, where the potential for the crystal field interaction at a specific site is given by:
\begin{equation}
V_{\mathrm{xtal}}({\bf r})=S_{\mathrm{xtal}}\sum^{N_{\mathrm{ions}}}_{m=1} \frac{Q_m} {|{\bf r}- {\bf R}_m|}.
\end{equation}
Fits to spectra on the meV energy scale are insensitive to the electrostatic and spin-orbit coupling, and hence only the crystal field scaling factor, $S_{\mathrm{xtal}}$ can be  refined.  The values found for \dy\ are $S_{\mathrm{xtal}}^{4c1}=0.35$ and $S_{\mathrm{xtal}}^{4c2}=0.53$, whereas for the Ho analogue we find $S_{\mathrm{xtal}}^{4c1}=0.62$ and $S_{\mathrm{xtal}}^{4c2}=0.70$. As $S_{\mathrm{xtal}}$ is dependent on the orbital overlap or covalency of the atom, the less-than-one scaling values may mean a significant $f$-orbital contraction is taking place, and/or the charges allocated to ions (Dy$^{3+}$, Ho$^{3+}$, Sr$^{2+}$, O$^{2-}$) are overestimated.  The excitations are described as Lorentzians, which are convoluted with a Gaussian to replicate the instrumental resolution \cite{Rosenkranz:2000p7896}.  The peak positions are calculated to the nearest meV. Although the range 20-70 meV was used for the fit Fig. \ref{fig:cef} shows that the low energy spectra is well described and the temperature dependence correctly reproduced.

The fits show that the first excited states of the Dy$^{3+}$ crystal field levels are found at 4 meV and 29 meV above the ground state doublet for the 4$c_1$ and the 4$c_2$ sites, respectively.  For Ho$^{3+}$, the splitting between the ground state and excited singlet levels is only about 1 meV for the 4$c_1$ site, and smaller than the computational accuracy of 0.3 meV for the 4$c_2$ site.  

\begin{table}
\begin{tabular}{|r|ccc|| r|ccc|}
\hline
& $\mu_\textrm{B}^\textrm{a}$ &  $\mu_\textrm{B}^\textrm{b}$&$\mu_\textrm{B}^\textrm{c}$ && $\mu_\textrm{B}^\textrm{a}$ &  $\mu_\textrm{B}^\textrm{b}$&$\mu_\textrm{B}^\textrm{c}$ \\
\hline
Dy$_{1_{cef}}$ & 0.7 & 1.5 & 7.7 & Dy$_{2_{cef}}$& 1.5 & 9.7 & 0.0\\
\hline
Ho$^{\text{pd}}_{1_{cef}}$ & 0.0 & 0.0 & 7.8 & Ho$^{\text{pd}}_{2_{cef}}$& 1.4 & 9.7 & 0.0\\
Ho$_{1_{exp}}$ & 0.000 & 0.000 & 6.080(3) & Ho$_{2_{exp}}$ & 0.000 & 7.740(3) & 0.130(3) \\
\hline
\end{tabular}
\caption{Moment sizes and orientations determined from crystal field calculations (cef), for the Dy ground state doublet and Ho pseudo-doublet (pd), and determined from neutron diffraction data (exp)\label{tab:mom}}
\end{table}

The magnetisation ellipsoid  can be determined from our model by calculation of the expected magnetic moment along the three crystallographic axes.  The ground state doublet is considered in the case of Dy. In the case of Ho it was assumed that the first excited state is equally populated.  The calculated moments, shown in Table \ref{tab:mom}, of both the Ho and Dy materials have an anisotropy, with the moment on the 4$c_1$ site lying predominantly along the $c$ axis and the moment on the  4$c_2$ lying along the $b$ axis.    Therefore, the fits indicate a strong Ising anisotropy in both materials. 

\begin{figure}
\includegraphics[width=8.5cm]{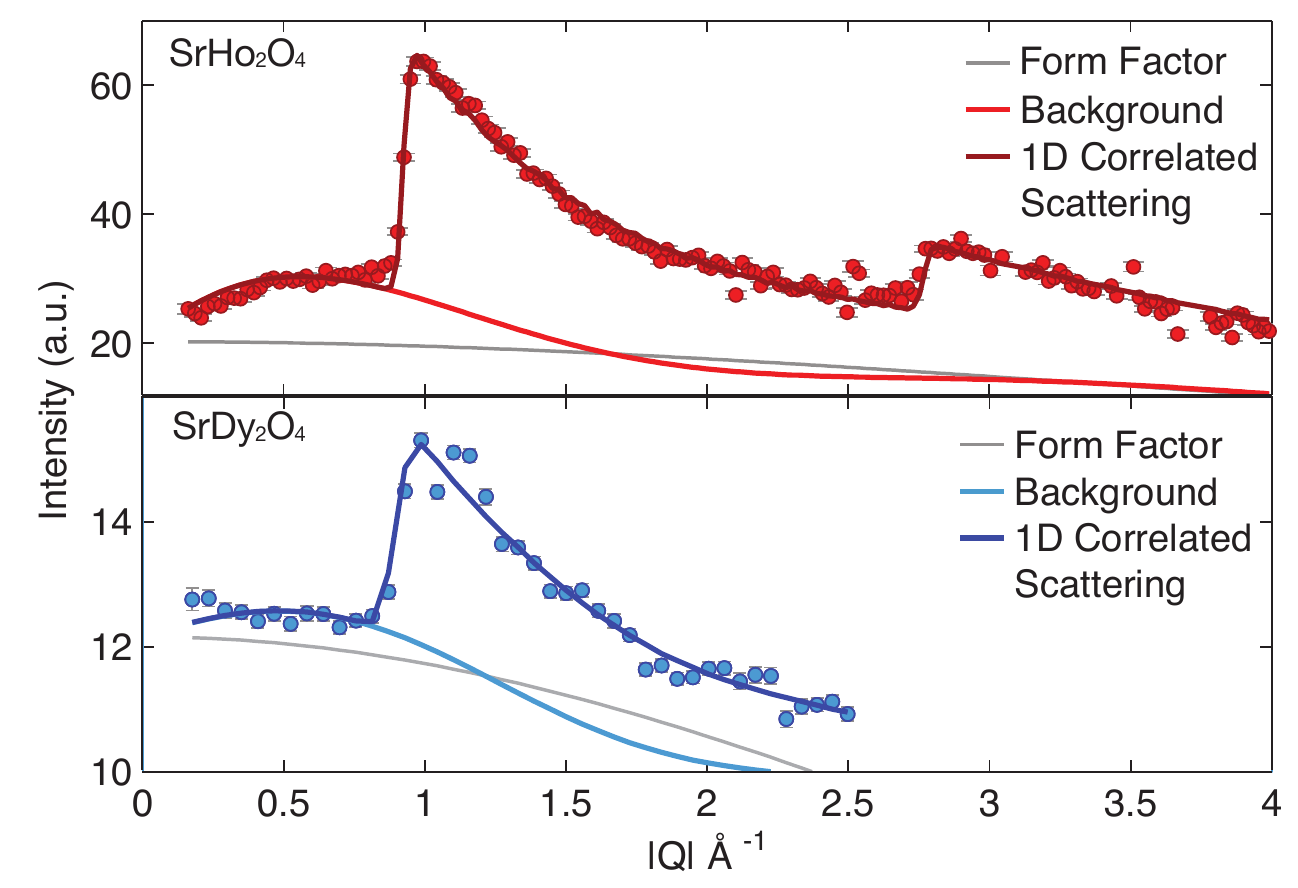}
\caption{\label{fig:neutronDiffuse}
(Color online) Diffuse neutron scattering observed in \ho\ and \dy\ at 0.8K and 0.05K, respectively.  The sharp features seen in the \ho ~scattering are indicative of 1D order.  The low temperature diffuse scattering in \dy\ has an intensity reduced by approximately a factor of 10 due to the strong absorption of the sample, the first maximum is again at {\bf Q}=[0 0 $n+0.5$].  }
\end{figure}

In order to experimentally determine the magnetic structure, neutron diffraction data were collected between T=50 mK and T=15 K for \dy\ and between T=50 mK and T=25 K for \ho\, using the diffractometer HRPT, PSI \cite{Fischer:2000p7956}.   The crystalline and magnetic structures were refined simultaneously with the Rietveld refinement method implemented in Fullprof \cite{RodriguezCarvajal:1993p7761} and the crystallographic structure was found to agree well with the published $Pnam$ structure ($\chi^2_{\textrm{Ho}}=1.869$ and $\chi^2_{\textrm{Dy}}=3.866$) \cite{Karunadasa:2005p3755}.    Prior to the condensation of 3D order \ho\ displayed a distinctive diffuse scattering pattern.  Furthermore,  \dy\ was found not to develop 3-dimensional magnetic order to T = 0.05 K, but instead presented a similar diffuse pattern.  The scattering, in both the samples, has a sharp feature, found at $Q_{\textrm{Ho}}$=0.95 \AA$^{-1}$  and $Q_{\textrm{Dy}}$=0.93 \AA$^{-1}$ close to the position of the respective [0 0 0.5] magnetic Bragg peaks, Fig. \ref{fig:neutronDiffuse}.  \ho\ has several of these features and the maxima correspond to wave-vector {\bf Q}=[0 0 $n+0.5$].  The features are nearly vertical on the low-$|Q|$ side and decay slowly on the high-$|Q|$ side, which is reminiscent the powder diffraction signature of 1D correlations \cite{JONES:1949p6859}.

In order to analyze the diffuse magnetic scattering, the high temperature nuclear scattering (with a Gaussian background subtraction) was subtracted from the  low temperature data  (T$_{\textrm{Ho}}$ = 0.8 K, T$_{\textrm{Dy}}$ =  0.05 K) to leave only the diffuse and paramagnetic scattering.  The data were then modelled by a combination of: a straight line, to capture the paramagnetic scattering; a damped sine wave, to describe the scattering from short-range correlations; and a powder averaged model of reciprocal space with {\bf Q}=[0 0 $n+0.5$]  planes of intensity,  to fit the 1D scattering.  The planes of scattering were assumed to be flat in the $hk$ plane, and the width in the $l$ direction could be controlled.  The different contribution from each component of the model were multiplied by the magnetic form factor and fitted with a scale factor.  

The best fit to the Ho data, ($\chi^2=7.02$), was given by a model with planes of intensity with a width of 0.035 r.l.u. in the $l$ direction, which gives a correlation length of 97.4 \AA\ for the magnetic correlations along the c axis (see Fig. \ref{fig:neutronDiffuse}).  The sine wave background was found to give a large amount of intensity centered $Q=0.567$ \AA$^{-1}$, close to the position of  {\bf Q}$= [ 0~ 0~ 1 ]$ Bragg peak, $Q=0.528$ \AA$^{-1}$, and indicates simultaneous short range 3D magnetic correlations, which are not in the $[h~ k~ 0.5]$ plane.   Assuming that the one-dimensional contribution to the diffuse scattering is from one site only, the integrated intensity of the scattering gives a moment size of 7.7(9)$\mu_B$, equivalent to the size of the ordered moment found from the refinement of the ordered phase.  

The best fit to the Dy data ($\chi^2=12.5$) was found with a scattering model that was 0.055 r.l.u. wide in the $l$ direction, which corresponds to a correlation length of 62 \AA\ along the $c$ axis.  Due to the substantial absorption of natural Dy only the region $Q<2.5$ \AA\ $^{-1}$ was included in the fit.  The sine wave background suggests a slight development of correlations with a maximum at $Q=0.582$ \AA$^{-1}$, but the intensity is much less significant than that found in \ho\, consistent with a much weaker tendency to long-range order in \dy\ compared to \ho.

The diffuse scattering indicates that 1D magnetic correlations are found in both of these materials, and that in \ho\ this coexists with short-range 3D correlations.  The 1D magnetic correlations can be understood to be dominated by interactions along the $c$ axis and the interchain-interaction to be extremely weak, to allow the 1D state.  Finally, because the scattering lies in the {\bf Q}=[$h~k~n+0.5$] planes the local order within the chains must be antiferromagnetic and similar to the magnetic order found below $T_N$ in \ho.

Below T$_N$ = 0.66 K the quasi-long-range ordered magnetic structure of \ho\ (the magnetic scattering is not resolution limited and single crystal measurements have shown the scattering is broad at all temperatures, \cite{Young:2013p7975}) was  found to have two types of magnetic order with different propagation vectors, ${\bf k }_0$=(0,0,0) and  ${\bf k}_{\frac{1}{2}}$=(0,0,$\frac{1}{2}$) ($\chi^2$=9.48).  These positions correspond to the two different components of the diffuse scattering.   The {\bf k}$_0$ order is described by the Pna$^\prime$m Shubnikov group with moments that are anti-ferromagnetically aligned along the $c$ axis.  The moment size is 6.080(3) $\mu_B$ on the first site and 0.130(3) $\mu_B$ on the second site, in good agreement with with Ref. \cite{Young:2012p7979}.

\begin{figure}
\includegraphics[width=6cm]{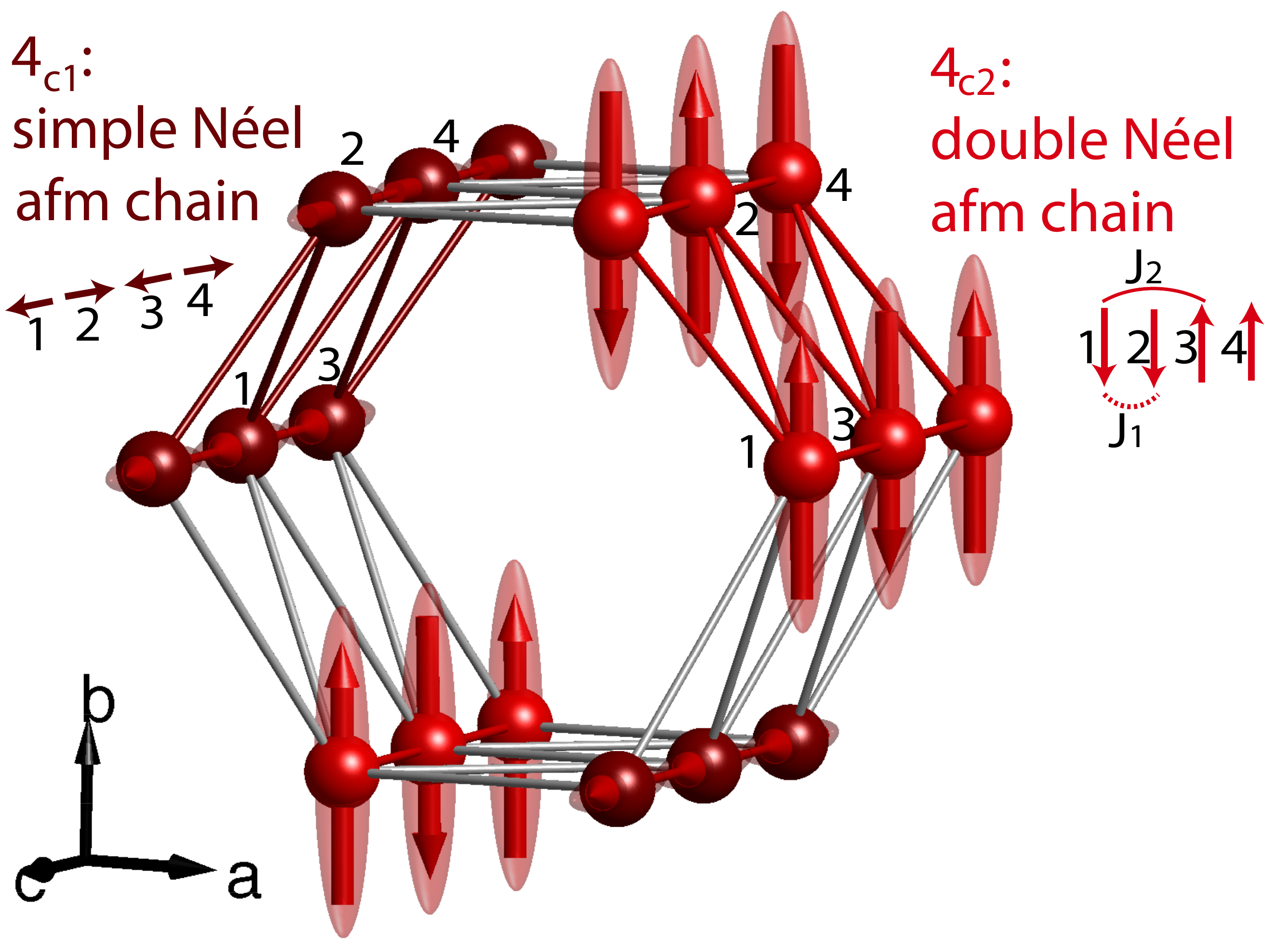}
\caption{\label{fig:magStructure}
(Color online) Refined magnetic structure of \ho.  The  moments on the Ho$_{c_1}$ sites are aligned with the $c$ axis and are ferromagnetic along the rungs and antiferromagnetic between the rungs.  The moments on the Ho$_{c_2}$ sites lie along the $b$ axis, and form an up-up-down-down structure along the chains. } 
\end{figure}

The {\bf k}$_{\frac{1}{2}}$=(0,0,$\frac{1}{2}$) component of the magnetic order, not determined in previous refinements, consists of moments with a magnitude of 7.740(3)$\mu_B$~along the $b$ axis,  which propagate in an up-up-down-down configuration along the chain.  Each type of order is predominantly associated with only one of the crystallographic sites, which we can uniquely assign due to the local anisotropy found from the crystal field calculations: the {\bf k}$_0$ order condenses on the  4c$_1$  site and {\bf k}$_{\frac{1}{2}}$ order on the  4c$_2$  site (Fig. \ref{fig:magStructure}).

The up-up-down-down, {\bf k}$_{\frac{1}{2}}$ magnetic structure, is equivalent to that expected for a finite Ising chain when $J_2/ J_1>1/2$ \cite{HeidrichMeisner:2007p7967, Igarashi:1989p7867}.  Our refinement of the {\bf k}$_{\frac{1}{2}}$ magnetic structure, combined with the evidence for the 1D nature of the correlations on the 4c$_2$ site and the strong anisotropy, determined from the crystal field, all indicate that the moments on the 4c$_2$ chain of Ho atoms, and by implication the 4c$_2$ Dy atoms are described by an Ising $J_1-J_2$ chain.

Recent magnetisation measurements find a $\frac{1}{3}$ plateau  when a magnetic field is applied along the $b$ axis, \cite{Hayes:2012p6555} i.e. parallel to the spins within the  4c$_2$ up-up-down-down chains, and perpendicular to the spins on the  4c$_1$ sites.   The plateau is predicted within the Ising $J_1-J_2$ model and the observation strongly supports the conclusion that \ho\ and \dy\ can be described as a classical Ising $J_1-J_2$ chain.  The plateau corresponds to an up-up-down magnetic structure, which precedes a ferromagnetic phase in the $S=\frac{1}{2}$ model, or an incommensurate phase followed by a ferromagnetic phase in the classical model \cite{Okunishi:2003p7870,HeidrichMeisner:2007p7967}.

The cross-over to the plateau region in \ho\ and \dy\  is H = 0.59 T and H = 0.16 T and to the saturated phase at H = 1.2 T and H = 2.03 T, respectively  \cite{Hayes:2012p6555}, which gives $J_{2}/J_{1Ho}   =   1.95$ and $J_{2}/J_{1Dy}    =   0.628$.  The nearest and next-nearest neighbor interactions are thus strongly competing, locating the magnetism of the zig-zag chains in the limit of $J_2/J_1>\frac{1}{2}$.  The findings strongly suggest that \dy\ is a model system for the classical Ising $J_1-J_2$ chain close to its quantum critical point and the emergence of the 1D magnetic correlations is the result of strong spin anisotropies and of frustrated interchain interaction. 

The spin anisotropy leads to the emergent 1D magnetic physics in these materials as the magnetic moments on neighboring chains are, in the most part, orientated perpendicular to each other.   Symmetric exchange, therefore, can not induce interchain correlations.  Furthermore, our symmetry analysis of the allowed Dzyaloshinskiy-Moriya (DM) interactions shows that they favor interchain order that is incompatible with the dominant spin correlations in the zig-zag chain.    The materials can, therefore, be understood as having a  dimensional reduction due to the spin anisotropy.

It is also clear why \ho\ has a higher tendency to magnetic order than \dy : The $J_2/J_1$ is much larger for \ho\ than for \dy\, so that less ground state fluctuations can be expected in \ho.  In fact, J2/J1 for SrDy2O4 puts this material close to the quantum critical point of the J1-J2 chain.  In addition, the lesser degree of spin anisotropy in \dy\ may enhance the phase space for fluctuations that can survive to much lower temperatures, and this may also contribute to the stabilization of a spin liquid ground state.

In summary, we determined the spin anisotropies in \ho\ and \dy\, by a fit of the crystal field excitations, and revealed the defining features of the magnetic Hamiltonian of these two materials. We conclude that the spin physics in these materials is dominated by emergent 1D correlations and are described by the $J_1-J_2$ Ising chain model. This now makes it possible to study this important model system, including its excitations, in its classical limit in more detail.

\section{Acknowledgements}
A. D. Bianchi received support from the Natural Sciences and Engineering Research Council of Canada (Canada), Fonds Qu\'{e}b\'{e}cois de la Recherche sur la Nature et les Technologies (Qu\'{e}bec), and the Canada Research Chair Foundation.  A Uldry acknowledges SNCF grant 20021-129970. We thank Michel Gingras for discussing the crystal fields, Tom Fennell for critical reading and Markus Zolliker for sample environment support.

\bibliography{RefDyHo}

\begin{thebibliography}{27}
\expandafter\ifx\csname natexlab\endcsname\relax\def\natexlab#1{#1}\fi
\expandafter\ifx\csname bibnamefont\endcsname\relax
  \def\bibnamefont#1{#1}\fi
\expandafter\ifx\csname bibfnamefont\endcsname\relax
  \def\bibfnamefont#1{#1}\fi
\expandafter\ifx\csname citenamefont\endcsname\relax
  \def\citenamefont#1{#1}\fi
\expandafter\ifx\csname url\endcsname\relax
  \def\url#1{\texttt{#1}}\fi
\expandafter\ifx\csname urlprefix\endcsname\relax\def\urlprefix{URL }\fi
\providecommand{\bibinfo}[2]{#2}
\providecommand{\eprint}[2][]{\url{#2}}

\bibitem[{\citenamefont{Gardner et~al.}(2010)\citenamefont{Gardner, Gingras,
  and Greedan}}]{Gardner:2010p2400}
\bibinfo{author}{\bibfnamefont{J.~S.} \bibnamefont{Gardner}},
  \bibinfo{author}{\bibfnamefont{M.~J.~P.} \bibnamefont{Gingras}},
  \bibnamefont{and} \bibinfo{author}{\bibfnamefont{J.~E.}
  \bibnamefont{Greedan}}, \bibinfo{journal}{Rev. Mod. Phys}
  \textbf{\bibinfo{volume}{82}}, \bibinfo{pages}{53} (\bibinfo{year}{2010}).

\bibitem[{\citenamefont{Greedan}(2006)}]{Greedan:2006p670}
\bibinfo{author}{\bibfnamefont{J.~E.} \bibnamefont{Greedan}},
  \bibinfo{journal}{J. Alloy Compd} \textbf{\bibinfo{volume}{408}},
  \bibinfo{pages}{444} (\bibinfo{year}{2006}).

\bibitem[{\citenamefont{Karunadasa et~al.}(2005)\citenamefont{Karunadasa,
  Huang, Ueland, Lynn, Schiffer, Regan, and Cava}}]{Karunadasa:2005p3755}
\bibinfo{author}{\bibfnamefont{H.}~\bibnamefont{Karunadasa}},
  \bibinfo{author}{\bibfnamefont{Q.}~\bibnamefont{Huang}},
  \bibinfo{author}{\bibfnamefont{B.~G.} \bibnamefont{Ueland}},
  \bibinfo{author}{\bibfnamefont{J.~W.} \bibnamefont{Lynn}},
  \bibinfo{author}{\bibfnamefont{P.}~\bibnamefont{Schiffer}},
  \bibinfo{author}{\bibfnamefont{K.~A.} \bibnamefont{Regan}}, \bibnamefont{and}
  \bibinfo{author}{\bibfnamefont{R.~J.} \bibnamefont{Cava}},
  \bibinfo{journal}{Phys. Rev. B} \textbf{\bibinfo{volume}{71}},
  \bibinfo{pages}{144414} (\bibinfo{year}{2005}).

\bibitem[{\citenamefont{Petrenko et~al.}(2008)\citenamefont{Petrenko,
  Balakrishnan, Wilson, de~Brion, Suard, and Chapon}}]{Petrenko:2008p6833}
\bibinfo{author}{\bibfnamefont{O.~A.} \bibnamefont{Petrenko}},
  \bibinfo{author}{\bibfnamefont{G.}~\bibnamefont{Balakrishnan}},
  \bibinfo{author}{\bibfnamefont{N.~R.} \bibnamefont{Wilson}},
  \bibinfo{author}{\bibfnamefont{S.}~\bibnamefont{de~Brion}},
  \bibinfo{author}{\bibfnamefont{E.}~\bibnamefont{Suard}}, \bibnamefont{and}
  \bibinfo{author}{\bibfnamefont{L.~C.} \bibnamefont{Chapon}},
  \bibinfo{journal}{Phys. Rev. B} \textbf{\bibinfo{volume}{78}},
  \bibinfo{pages}{184410} (\bibinfo{year}{2008}).

\bibitem[{\citenamefont{Hayes et~al.}(2012)\citenamefont{Hayes, Young,
  Balakrishnan, and Petrenko}}]{Hayes:2012p6555}
\bibinfo{author}{\bibfnamefont{T.~J.} \bibnamefont{Hayes}},
  \bibinfo{author}{\bibfnamefont{O.}~\bibnamefont{Young}},
  \bibinfo{author}{\bibfnamefont{G.}~\bibnamefont{Balakrishnan}},
  \bibnamefont{and} \bibinfo{author}{\bibfnamefont{O.~A.}
  \bibnamefont{Petrenko}}, \bibinfo{journal}{J. Phys. Soc. Jpn}
  \textbf{\bibinfo{volume}{81}}, \bibinfo{pages}{024708}
  (\bibinfo{year}{2012}).

\bibitem[{\citenamefont{Quintero-Castro
  et~al.}(2012)\citenamefont{Quintero-Castro, Lake, Reehuis, Niazi, Ryll,
  Islam, Fennell, Kimber, Klemke, Ollivier et~al.}}]{QuinteroCastro:2012p7953}
\bibinfo{author}{\bibfnamefont{D.~L.} \bibnamefont{Quintero-Castro}},
  \bibinfo{author}{\bibfnamefont{B.}~\bibnamefont{Lake}},
  \bibinfo{author}{\bibfnamefont{M.}~\bibnamefont{Reehuis}},
  \bibinfo{author}{\bibfnamefont{A.}~\bibnamefont{Niazi}},
  \bibinfo{author}{\bibfnamefont{H.}~\bibnamefont{Ryll}},
  \bibinfo{author}{\bibfnamefont{A.~T. M.~N.} \bibnamefont{Islam}},
  \bibinfo{author}{\bibfnamefont{T.}~\bibnamefont{Fennell}},
  \bibinfo{author}{\bibfnamefont{S.~A.~J.} \bibnamefont{Kimber}},
  \bibinfo{author}{\bibfnamefont{B.}~\bibnamefont{Klemke}},
  \bibinfo{author}{\bibfnamefont{J.}~\bibnamefont{Ollivier}},
  \bibnamefont{et~al.}, \bibinfo{journal}{Phys Rev B}
  \textbf{\bibinfo{volume}{86}}, \bibinfo{pages}{064203}
  (\bibinfo{year}{2012}).

\bibitem[{\citenamefont{Young et~al.}(2013)\citenamefont{Young, Wildes, Manuel,
  Ouladdiaf, Khalyavin, Balakrishnan, and Petrenko}}]{Young:2013p7975}
\bibinfo{author}{\bibfnamefont{O.}~\bibnamefont{Young}},
  \bibinfo{author}{\bibfnamefont{A.~R.} \bibnamefont{Wildes}},
  \bibinfo{author}{\bibfnamefont{P.}~\bibnamefont{Manuel}},
  \bibinfo{author}{\bibfnamefont{B.}~\bibnamefont{Ouladdiaf}},
  \bibinfo{author}{\bibfnamefont{D.~D.} \bibnamefont{Khalyavin}},
  \bibinfo{author}{\bibfnamefont{G.}~\bibnamefont{Balakrishnan}},
  \bibnamefont{and} \bibinfo{author}{\bibfnamefont{O.~A.}
  \bibnamefont{Petrenko}}, \bibinfo{journal}{Physical Review B}
  \textbf{\bibinfo{volume}{88}}, \bibinfo{pages}{024411}
  (\bibinfo{year}{2013}).

\bibitem[{\citenamefont{Majumdar and Ghosh}(1969)}]{Majumdar:1969p7877}
\bibinfo{author}{\bibfnamefont{C.}~\bibnamefont{Majumdar}} \bibnamefont{and}
  \bibinfo{author}{\bibfnamefont{D.}~\bibnamefont{Ghosh}},
  \bibinfo{journal}{Journal of Mathematical Physics}
  \textbf{\bibinfo{volume}{10}}, \bibinfo{pages}{1388} (\bibinfo{year}{1969}).

\bibitem[{\citenamefont{Morita and Horiguchi}(1972)}]{Morita:1972p7901}
\bibinfo{author}{\bibfnamefont{T.}~\bibnamefont{Morita}} \bibnamefont{and}
  \bibinfo{author}{\bibfnamefont{T.}~\bibnamefont{Horiguchi}},
  \bibinfo{journal}{Physics Letters A} \textbf{\bibinfo{volume}{38}},
  \bibinfo{pages}{223} (\bibinfo{year}{1972}).

\bibitem[{\citenamefont{Heidrich-Meisner
  et~al.}(2007)\citenamefont{Heidrich-Meisner, Sergienko, Feiguin, and
  Dagotto}}]{HeidrichMeisner:2007p7967}
\bibinfo{author}{\bibfnamefont{F.}~\bibnamefont{Heidrich-Meisner}},
  \bibinfo{author}{\bibfnamefont{I.~A.} \bibnamefont{Sergienko}},
  \bibinfo{author}{\bibfnamefont{A.~E.} \bibnamefont{Feiguin}},
  \bibnamefont{and} \bibinfo{author}{\bibfnamefont{E.~R.}
  \bibnamefont{Dagotto}}, \bibinfo{journal}{Phys. Rev. B}
  \textbf{\bibinfo{volume}{75}}, \bibinfo{pages}{064413}
  (\bibinfo{year}{2007}).

\bibitem[{\citenamefont{Igarashi and Tonegawa}(1989)}]{Igarashi:1989p7867}
\bibinfo{author}{\bibfnamefont{J.~I.} \bibnamefont{Igarashi}} \bibnamefont{and}
  \bibinfo{author}{\bibfnamefont{T.}~\bibnamefont{Tonegawa}},
  \bibinfo{journal}{Phys. Rev. B} \textbf{\bibinfo{volume}{40}},
  \bibinfo{pages}{756} (\bibinfo{year}{1989}).

\bibitem[{\citenamefont{Okunishi and Tonegawa}(2003)}]{Okunishi:2003p7870}
\bibinfo{author}{\bibfnamefont{K.}~\bibnamefont{Okunishi}} \bibnamefont{and}
  \bibinfo{author}{\bibfnamefont{T.}~\bibnamefont{Tonegawa}},
  \bibinfo{journal}{Phys Rev B} \textbf{\bibinfo{volume}{68}},
  \bibinfo{pages}{224422} (\bibinfo{year}{2003}).

\bibitem[{\citenamefont{Castilla et~al.}(1995)\citenamefont{Castilla,
  Chakravarty, and Emery}}]{Castilla:1995p7939}
\bibinfo{author}{\bibfnamefont{G.}~\bibnamefont{Castilla}},
  \bibinfo{author}{\bibfnamefont{S.}~\bibnamefont{Chakravarty}},
  \bibnamefont{and} \bibinfo{author}{\bibfnamefont{V.~J.} \bibnamefont{Emery}},
  \bibinfo{journal}{Physical review letters} \textbf{\bibinfo{volume}{75}},
  \bibinfo{pages}{1823} (\bibinfo{year}{1995}).

\bibitem[{\citenamefont{Enderle et~al.}(2005)\citenamefont{Enderle, Mukherjee,
  F{\aa}k, Kremer, Broto, Rosner, Drechsler, Richter, Malek, and
  Prokofiev}}]{Enderle:2005p7932}
\bibinfo{author}{\bibfnamefont{M.}~\bibnamefont{Enderle}},
  \bibinfo{author}{\bibfnamefont{C.}~\bibnamefont{Mukherjee}},
  \bibinfo{author}{\bibfnamefont{B.}~\bibnamefont{F{\aa}k}},
  \bibinfo{author}{\bibfnamefont{R.~K.} \bibnamefont{Kremer}},
  \bibinfo{author}{\bibfnamefont{J.~M.} \bibnamefont{Broto}},
  \bibinfo{author}{\bibfnamefont{H.}~\bibnamefont{Rosner}},
  \bibinfo{author}{\bibfnamefont{S.~L.} \bibnamefont{Drechsler}},
  \bibinfo{author}{\bibfnamefont{J.}~\bibnamefont{Richter}},
  \bibinfo{author}{\bibfnamefont{J.}~\bibnamefont{Malek}}, \bibnamefont{and}
  \bibinfo{author}{\bibfnamefont{A.}~\bibnamefont{Prokofiev}},
  \bibinfo{journal}{EPL (Europhysics Letters)} \textbf{\bibinfo{volume}{70}},
  \bibinfo{pages}{237} (\bibinfo{year}{2005}).

\bibitem[{\citenamefont{Drechsler et~al.}(2007)\citenamefont{Drechsler,
  Volkova, Vasiliev, Tristan, Richter, Schmitt, Rosner, M{\'a}lek, Klingeler,
  Zvyagin et~al.}}]{Drechsler:2007p7935}
\bibinfo{author}{\bibfnamefont{S.-L.} \bibnamefont{Drechsler}},
  \bibinfo{author}{\bibfnamefont{O.}~\bibnamefont{Volkova}},
  \bibinfo{author}{\bibfnamefont{A.}~\bibnamefont{Vasiliev}},
  \bibinfo{author}{\bibfnamefont{N.}~\bibnamefont{Tristan}},
  \bibinfo{author}{\bibfnamefont{J.}~\bibnamefont{Richter}},
  \bibinfo{author}{\bibfnamefont{M.}~\bibnamefont{Schmitt}},
  \bibinfo{author}{\bibfnamefont{H.}~\bibnamefont{Rosner}},
  \bibinfo{author}{\bibfnamefont{J.}~\bibnamefont{M{\'a}lek}},
  \bibinfo{author}{\bibfnamefont{R.}~\bibnamefont{Klingeler}},
  \bibinfo{author}{\bibfnamefont{A.}~\bibnamefont{Zvyagin}},
  \bibnamefont{et~al.}, \bibinfo{journal}{Physical review letters}
  \textbf{\bibinfo{volume}{98}}, \bibinfo{pages}{077202}
  (\bibinfo{year}{2007}).

\bibitem[{\citenamefont{Goff et~al.}(1995)\citenamefont{Goff, Tennant, and
  Nagler}}]{Goff:1995p7928}
\bibinfo{author}{\bibfnamefont{J.~P.} \bibnamefont{Goff}},
  \bibinfo{author}{\bibfnamefont{D.~A.} \bibnamefont{Tennant}},
  \bibnamefont{and} \bibinfo{author}{\bibfnamefont{S.~E.}
  \bibnamefont{Nagler}}, \bibinfo{journal}{Physical Review B}
  \textbf{\bibinfo{volume}{52}}, \bibinfo{pages}{15992} (\bibinfo{year}{1995}).

\bibitem[{\citenamefont{Wolf}(2000)}]{Wolf:2000p7944}
\bibinfo{author}{\bibfnamefont{W.~P.} \bibnamefont{Wolf}},
  \bibinfo{journal}{Brazilian Journal of Physics}
  \textbf{\bibinfo{volume}{30}}, \bibinfo{pages}{794} (\bibinfo{year}{2000}).

\bibitem[{\citenamefont{Coldea et~al.}(2010)\citenamefont{Coldea, Tennant,
  Wheeler, Wawrzynska, Prabhakaran, Telling, Habicht, Smeibidl, and
  Kiefer}}]{Coldea:2010p7931}
\bibinfo{author}{\bibfnamefont{R.}~\bibnamefont{Coldea}},
  \bibinfo{author}{\bibfnamefont{D.~A.} \bibnamefont{Tennant}},
  \bibinfo{author}{\bibfnamefont{E.~M.} \bibnamefont{Wheeler}},
  \bibinfo{author}{\bibfnamefont{E.}~\bibnamefont{Wawrzynska}},
  \bibinfo{author}{\bibfnamefont{D.}~\bibnamefont{Prabhakaran}},
  \bibinfo{author}{\bibfnamefont{M.}~\bibnamefont{Telling}},
  \bibinfo{author}{\bibfnamefont{K.}~\bibnamefont{Habicht}},
  \bibinfo{author}{\bibfnamefont{P.}~\bibnamefont{Smeibidl}}, \bibnamefont{and}
  \bibinfo{author}{\bibfnamefont{K.}~\bibnamefont{Kiefer}},
  \bibinfo{journal}{Science} \textbf{\bibinfo{volume}{327}},
  \bibinfo{pages}{177} (\bibinfo{year}{2010}).

\bibitem[{\citenamefont{Matsuda and Katsumata}(1995)}]{Matsuda:1995p7933}
\bibinfo{author}{\bibfnamefont{M.}~\bibnamefont{Matsuda}} \bibnamefont{and}
  \bibinfo{author}{\bibfnamefont{K.}~\bibnamefont{Katsumata}},
  \bibinfo{journal}{J Magn Magn Mater} \textbf{\bibinfo{volume}{140}},
  \bibinfo{pages}{1671} (\bibinfo{year}{1995}).

\bibitem[{\citenamefont{Walter}(1984)}]{Walter:1984p7623}
\bibinfo{author}{\bibfnamefont{U.}~\bibnamefont{Walter}}, \bibinfo{journal}{J.
  Phys. Chem. Solids} \textbf{\bibinfo{volume}{45}}, \bibinfo{pages}{401}
  (\bibinfo{year}{1984}).

\bibitem[{\citenamefont{Cheffings et~al.}(2013)\citenamefont{Cheffings, Lees,
  Balakrishnan, and Petrenko}}]{Cheffings:2013p7898}
\bibinfo{author}{\bibfnamefont{T.}~\bibnamefont{Cheffings}},
  \bibinfo{author}{\bibfnamefont{M.}~\bibnamefont{Lees}},
  \bibinfo{author}{\bibfnamefont{G.}~\bibnamefont{Balakrishnan}},
  \bibnamefont{and} \bibinfo{author}{\bibfnamefont{O.}~\bibnamefont{Petrenko}},
  \bibinfo{journal}{Journal of Physics: Condensed Matter}
  \textbf{\bibinfo{volume}{25}}, \bibinfo{pages}{256001}
  (\bibinfo{year}{2013}).

\bibitem[{\citenamefont{Uldry et~al.}(2012)\citenamefont{Uldry, Vernay, and
  Delley}}]{Uldry:2012p7902}
\bibinfo{author}{\bibfnamefont{A.}~\bibnamefont{Uldry}},
  \bibinfo{author}{\bibfnamefont{F.}~\bibnamefont{Vernay}}, \bibnamefont{and}
  \bibinfo{author}{\bibfnamefont{B.}~\bibnamefont{Delley}},
  \bibinfo{journal}{Physical Review B} \textbf{\bibinfo{volume}{85}},
  \bibinfo{pages}{125133} (\bibinfo{year}{2012}).

\bibitem[{\citenamefont{Rosenkranz et~al.}(2000)\citenamefont{Rosenkranz,
  Ramirez, Hayashi, Cava, Siddharthan, and Shastry}}]{Rosenkranz:2000p7896}
\bibinfo{author}{\bibfnamefont{S.}~\bibnamefont{Rosenkranz}},
  \bibinfo{author}{\bibfnamefont{A.~P.} \bibnamefont{Ramirez}},
  \bibinfo{author}{\bibfnamefont{A.}~\bibnamefont{Hayashi}},
  \bibinfo{author}{\bibfnamefont{R.~J.} \bibnamefont{Cava}},
  \bibinfo{author}{\bibfnamefont{R.}~\bibnamefont{Siddharthan}},
  \bibnamefont{and} \bibinfo{author}{\bibfnamefont{B.~S.}
  \bibnamefont{Shastry}}, \bibinfo{journal}{Journal Of Applied Physics}
  \textbf{\bibinfo{volume}{87}}, \bibinfo{pages}{5914} (\bibinfo{year}{2000}).

\bibitem[{\citenamefont{Fischer et~al.}(2000)\citenamefont{Fischer, Frey, Koch,
  K{\"o}nnecke, Pomjakushin, Schefer, Thut, Schlumpf, B{\"u}rge, and
  Greuter}}]{Fischer:2000p7956}
\bibinfo{author}{\bibfnamefont{P.}~\bibnamefont{Fischer}},
  \bibinfo{author}{\bibfnamefont{G.}~\bibnamefont{Frey}},
  \bibinfo{author}{\bibfnamefont{M.}~\bibnamefont{Koch}},
  \bibinfo{author}{\bibfnamefont{M.}~\bibnamefont{K{\"o}nnecke}},
  \bibinfo{author}{\bibfnamefont{V.}~\bibnamefont{Pomjakushin}},
  \bibinfo{author}{\bibfnamefont{J.}~\bibnamefont{Schefer}},
  \bibinfo{author}{\bibfnamefont{R.}~\bibnamefont{Thut}},
  \bibinfo{author}{\bibfnamefont{N.}~\bibnamefont{Schlumpf}},
  \bibinfo{author}{\bibfnamefont{R.}~\bibnamefont{B{\"u}rge}},
  \bibnamefont{and} \bibinfo{author}{\bibfnamefont{U.}~\bibnamefont{Greuter}},
  \bibinfo{journal}{Physica B: Condensed Matter}
  \textbf{\bibinfo{volume}{276}}, \bibinfo{pages}{146} (\bibinfo{year}{2000}).

\bibitem[{\citenamefont{Rodr{\'\i}guez-Carvajal}(1993)}]{RodriguezCarvajal:199%
3p7761}
\bibinfo{author}{\bibfnamefont{J.}~\bibnamefont{Rodr{\'\i}guez-Carvajal}},
  \bibinfo{journal}{Physica B: Condensed Matter}
  \textbf{\bibinfo{volume}{192}}, \bibinfo{pages}{55} (\bibinfo{year}{1993}).

\bibitem[{\citenamefont{Jones}(1949)}]{JONES:1949p6859}
\bibinfo{author}{\bibfnamefont{R.}~\bibnamefont{Jones}}, \bibinfo{journal}{Acta
  Crystallogr.} \textbf{\bibinfo{volume}{2}}, \bibinfo{pages}{252}
  (\bibinfo{year}{1949}).

\bibitem[{\citenamefont{Young et~al.}(2012)\citenamefont{Young, Chapon, and
  Petrenko}}]{Young:2012p7979}
\bibinfo{author}{\bibfnamefont{O.}~\bibnamefont{Young}},
  \bibinfo{author}{\bibfnamefont{L.}~\bibnamefont{Chapon}}, \bibnamefont{and}
  \bibinfo{author}{\bibfnamefont{O.}~\bibnamefont{Petrenko}},
  \textbf{\bibinfo{volume}{391}}, \bibinfo{pages}{012081}
  (\bibinfo{year}{2012}).

\end{thebibliography}

\end{document}